\documentclass[pre,superscriptaddress,showpacs,nofootinbib,
  floatfix,preprint,preprintnumbers]{revtex4-1}

\usepackage[T1]{fontenc}

\usepackage{amsfonts}
\usepackage{amsmath}
\usepackage{amssymb}
\usepackage[pdftex]{graphicx}
\usepackage{epstopdf}
\usepackage{hyperref}
\usepackage{bm}
\usepackage{upgreek}
\usepackage{subcaption}
\usepackage{lmodern}
\usepackage{natbib}
\hypersetup{colorlinks=true, citecolor=blue,
  pagecolor=blue,
  urlcolor=black,
  pdfcreator={pdflatex},
}

\begin{document}

\title{Waterlike anomalies in hard core-soft shell nanoparticles using a effective potential approach: pinned vs adsorbed polymers}

\author{Murilo S. Marques}
\email{murilo.sodre@ufob.edu.br}
\affiliation{Centro das Ci\^encias Exatas e das Tecnologias, Universidade Federal do Oeste da Bahia, Barreiras - BA}
\affiliation {Instituto de F\'isica, Universidade Federal do Rio Grande do Sul, , Caixa Postal 15051, CEP 91501-970, Porto Alegre, Rio Grande do Sul, Brazil}

\author{Thiago P. Nogueira}
\affiliation {Departamento de F\'isica, Instituto de F\'isica e Matem\'atica, Universidade Federal de Pelotas, Caixa Postal 354, 96010-900, Pelotas, Rio Grande do Sul, Brazil}

\author{Marcia C.  Barbosa}
\affiliation {Instituto de F\'isica, Universidade Federal do Rio Grande do Sul, , Caixa Postal 15051, CEP 91501-970, Porto Alegre, Rio Grande do Sul, Brazil}

\author{Jos\'e Rafael Bordin}
\email{jrbordin@ufpel.edu.br}
\affiliation {Departamento de F\'isica, Instituto de F\'isica e Matem\'atica, Universidade Federal de Pelotas, Caixa Postal 354, 96010-900, Pelotas, Rio Grande do Sul, Brazil}

\begin{abstract}

 In this work, a two dimensional system of polymer grafted nanoparticles is analyzed using large-scale Langevin Dymanics simulations. Effective core-softened potentials were obtained for two cases: one where the polymers are free to rotate around the nanoparticle core and a second where the polymers are fixed, with a $45^\circ$ angle between them. The use of effective core-softened potentials allow us to explore the complete system phase space. In this way, the $PT$, $T\rho$ and $P\rho$ phase diagrams for each potential were obtained, with all fluid and solid phases. The phase boundaries were defined analyzing the specific heat at constant pressure, the system mean square displacement, the radial distribution function and  the discontinuities in the density-pressure phase diagram. Also, due the competition in the system  we have observed the presence of waterlike anomalies, such as the temperature of maximum density - in addition with a tendency of the TMD to move to lower temperatures (negative slope)- and the diffusion anomaly. It was observed different morphologies (stripes, honeycomb, amorphous) for each nanoparticle. We observed that for the fixed polymers case the waterlike anomalies are originated by the competition between the potential characteristic length scales, while for the free to rotate case the anomalies arises due a smaller region of stability in the phase diagram and no competition between the scales was observed.
 
\end{abstract}

\clearpage
\maketitle

\section{Introduction}

Coarse-grained (CG) representations of macromolecular liquids have gained widespread  interest because of their ability to represent large-scale  properties of systems that cannot be investigated by atomistic scale simulations because of their large size and long  timescales~\cite{Clark2013,Renevey2017}. 

Among coarse-grained models, core-softened (CS) potentials (characterized by having two preferred particle-particle separations) 
have been attracting attention due to their connections with the anomalous behavior of liquid systems including water.  
They show a variety of shapes: they can be ramp-like~\cite{Jagla1999} or continuous shoulder-like~\cite{Evy2011,Formin2011,Ney2009,jonatan2010}. 
Despite their simplicity, these models originate from the desire of constructing a simple two-body isotropic potential capable of describing
the complicated features of systems interacting via anisotropic potentials~\cite{birnbaum2013,kaplan2006}, and are able to  reproduce waterlike 
anomalies in qualitative way if competition exists between the two characteristic  distances~\cite{Galo2016,Vilaseca2011}. If the energy penalty 
to the particle moves from one scale to another is higher than the particle kinetic energy, then the particle will get trapped in one length scale, 
and there will be  no competition. As a consequence, there will be no anomalous behavior. This procedure generates models that are  analytically 
tractable and computationally less expensive than the atomistic models. Moreover, they are lead to conclusions that are more universal and are 
related to families of atomistic systems~\cite{Tsuchiya1991}.

The study of chemical building blocks as amphiphilic molecules, colloids and nanoparticles have attracted much attention in soft matter physical chemistry in recent years due their properties of self-assembly \cite{Sacanna2013,Pham2015,Preisler2013, Kumar17}. When in water solution, these large molecules agglomerate. In order to circumvent this phase separation, one the most important practical methods for stabilizing colloids is by coating the particle with a polymers layers~\cite{jones2002,witten2004}. These polymer-grafted nanoparticles (GNPs), composed of an inorganic core and a grafted layer of polymer chains  possess new intriguing   electrical  conductivity, optical and visco-elastic 
properties~\cite{Chremos2016,Chevigny2011,Ganesan2014,Sunday2015,Talapin2010} not present in the non coated system.   The generated self-assembled structures have applications in medicine, self-driven molecules,  catalysis, photonic crystals,  stable emulsions, biomolecules and self-healing materials~\cite{Zhang2015}.  Experiments~\cite{Lokupitiya2016} and  simulations~\cite{Lindquist2017,Schwanzer2016} showed that in the case of spherical colloids the mechanism behind the formation of these  distinct patterns is the presence of  competitive interactions. These  competing forces can appear from the  combination  of a short range attraction of the core  and a long-range repulsion~\cite{Shukla2008} of the grafted 
polymers~\cite{BoK15,Bok16b,Bordin16, Bordin18a,Bordin18b,Bordin19d}.

The objective of our work is to analyze the structural, thermodynamic and dynamic behavior of 2D polymer-grafted nanoparticle systems through effective potentials in light of molecular dynamics. Particularly, we are interested in how the specificity of the grafted polymers structure can affect the macroscopic morphology and dynamical behavior of these systems when absorbed in large surfaces or  when assembled in quasi 2D solid-liquid interfaces~\cite{Volk19}.

One of the characteristics of grafted nanoparticles is that by adding appropriated reactive  groups in their surface, it is possible to design new materials. In particular  polymers can be adsorbed to the nanoparticle core by fully or partially coating the surface. In the case of partially coated,  the polymers  are free to rotate at the nanoparticle surface. If the polymers are grafted to the surface by a reactive group or by the polymerization they can not rotate~\cite{Hore19}. 

Here we address the question of how the two types of attachments  affect the phase behavior of the nanoparticle solution.  We  adopt two complementary strategies. We model the systems using a CG approach in which the chemical interactions are represented by classical interactions.  Based in this CG model, we derive  effective potentials for the two cases:  polymers are pinned to a reactive group, and polymers grafted to the surface.  CG models have been used as a powerful tool to explore rather complex systems~\cite{Lafitte14, Song17}. The additional simplification of  using  effective potentials not only allow for exploring the complete pressure versus temperature phase diagram with a low computational cost but also is able to focus in the physical mechanism behind the different degrees of freedom of free and non free cases.

The remaining of the paper is organized as follows: in section 2 we discuss the model and details of the simulation. In section 3 the results and the  main discussions; in section 4, the conclusions are listed.

\section{The model and simulation details}

Here we employ two complementary approximations to describe the  polymer-grafted nanoparticles phase diagram: a coarse grained model and an effective core-softened potential.

\subsection{The Coarse-Grained (CG) Model}

We employ a two dimensional  coarse-grained model proposed in previous works~\cite{Ackora09, Hong12, Lafitte14} to describe 
a polymer-grafted nanoparticles interactions.  Each core-shell nanoparticle is composed of a central disk with diameter $\sigma_{core}$ 
and with 4 linear oligomer attached chains. Each chain consists of 3 beads with diameter $\sigma_{bead}$, connected by an harmonic bond
\begin{equation}
 U_{bond}(r_{ij}) = k\left( r_{ij} - \sigma_{bead} \right) 
\end{equation}
\noindent with $k = 5000$ in reduced LJ units. The bead-bead (bb) interaction is modeled by the standard Lennard Jones (LJ) potential
\begin{equation}
U_{bb}(r_{ij})=4\epsilon \left [\left (\frac{\sigma}{r_{ij}}  \right )^{12} - \left (\frac{\sigma}{r_{ij}}  \right )^{6} \right ] - U_{LJ}(r_{cut})
\end{equation}
\noindent where $r_{cut} = 2.5\sigma_{mon}$ and $\epsilon = \epsilon_{core}/9.0$. For the core-core (cc) interaction 
was used a 14-7 LJ potential, 
\begin{equation}
U_{cc}(r_{ij})=4\epsilon_c \left [\left (\frac{\sigma_{core}}{r_{ij}}  \right )^{14} - \left (\frac{\sigma_{core}}{r_{ij}}  \right )^{7} \right ] - U_{cc}(r_{cut})
\end{equation}
\noindent where $r_{cut} = 2.5\sigma_{core}$. 
Finally, the core-bead (cp) interaction if give by a 13.5-6.5 LJ potential,
\begin{equation}
U_{cp}(r_{ij})=4\epsilon_{cp} \left [\left (\frac{\sigma_{cp}}{r_{ij}}  \right )^{13.5} - \left (\frac{\sigma_{cp}}{r_{ij}}  
\right )^{6.5} \right ] - U_{cp}(r_{cut})
\end{equation}
\noindent where $\sigma_{cp}$ and $\epsilon_{cp}$ are obtained by the well-known Lorentz-Berthelot combining rules. 

\begin{figure}[htb]
\centering \includegraphics[width=12cm]{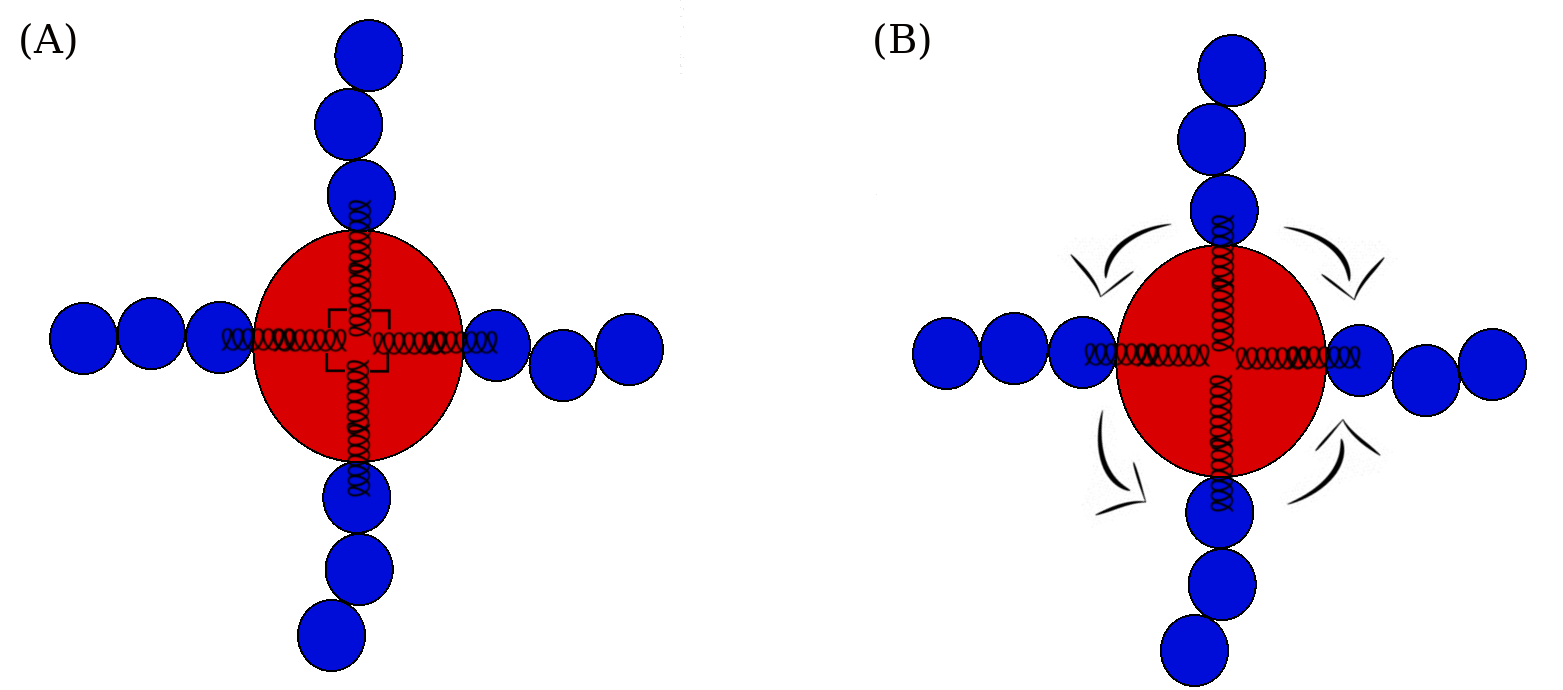}	
\caption{Schematic depiction of the CG nanoparticles. (A) Nanoparticles with a bend potential that prevents the polymers of slide along the core surface and (B) nanoparticles without bend potential whose polymers are free to rotate along the core surface.}
\label{fig:coarse-grained-particles}
\end{figure}

The first bead in the polymer chain is connected to the central core by a rigid bond~\cite{Ryc77}. Two cases of grafted NP 
were considered. In the first one the polymers are held fixed in the core surface with a  separation of 45$^o$ by the bend 
cosine square bond angle potential,
\begin{equation}
 U_{bend} = \frac{k_{bend}}{2}[cos(\phi) - \cos(\phi_0)]^2
\end{equation}
\noindent with $k_{bond}$ = 50 and $\phi_0 = \pi/4$. In the second case no bending potential was applied, and 
the  polymers are free to rotate around the central colloid. Both structures are illustrated in the figure~\ref{fig:coarse-grained-particles}.

In order to simulate a small silica core we use in this work 
$\sigma_{core} = 1.4$ nm and $\epsilon_{core}/k_b = 10179$ K,  as proposed by Lafitte and co-authors~\cite{Lafitte14}. 
The polymer beads have a diameter $\sigma_{bead} = 0.4$ nm and $\epsilon_{bead} = \epsilon_{core}/9.0$ and correspond to a 
ethoxy repeat unit~\cite{Hong12}. For simplicity, for now on this paper all physical quantities will be displayed in the standard LJ units. 
Distance, density of particles, time, pressure and temperature are given, respectively, by
\begin{equation}
r^*\equiv\frac{r}{\sigma_{core}}, \quad \rho^* \equiv \rho \sigma_{core}^3,\quad  t^* \equiv t\left ( \frac{\epsilon_{core}}{m\sigma_{core}^2} \right )^{1/2}, 
\quad p^*\equiv \frac{p\sigma_{core}^3 }{\epsilon_{core}}, \quad and \quad T^*\equiv \frac{k_BT}{\epsilon_{core}}.
\end{equation}

\subsection{The Effective Core-Softened Potential}

The effective core-softened (CS) potentials for the two polymer-grafted nanoparticles systems analyzed here were obtained  as follows. Langevin Dynamics simulations using the ESPResSo package~\cite{espresso1, espresso2} 
were performed for the coarse-grained models (fixed  and free beads). The two  systems were analyzed  
in the $NVT$ ensemble for  density $\rho^* = 0.25$ and  temperature $T^* = 0.5$. These values were chosen 
to ensure that the coarse-grained models were both in the fluid state.

 \begin{figure} [htb]
 \centering \includegraphics[width=12cm]{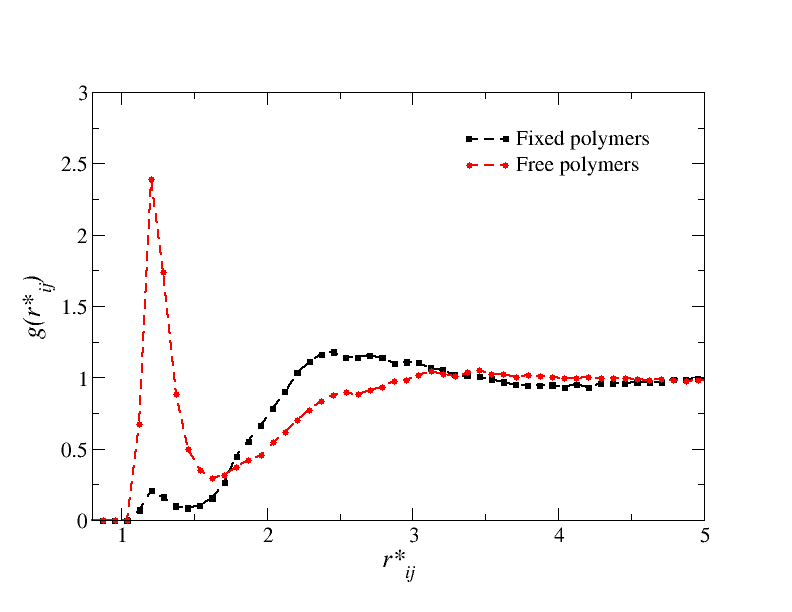}	
 \caption{Radial distribution functions employed to obtain the effective interaction potential between NPs with polymers fixed (black squares) or
 free (red circles) to rotate. Both RDFs were obtained at the density of $\rho^* = 0.25$ and  temperature $T^* = 0.5$.}
 \label{fig:coarse-grained-RDF}
 \end{figure}

Then,  the core-core radial distribution functions (RDF) for this state point for each fixed and free beads systems were computed as illustrated in  the figure~\ref{fig:coarse-grained-RDF}. As we can see, the RDFs indicates a significant difference in the length scales occupancy. For the case of fixed polymers, black curve in the figure~\ref{fig:coarse-grained-RDF}, it is harder for the cores remain close to each other. As consequence, this NP has a higher occupancy in the  second length scale (the polymer corona) and a smaller in the first length scale - the hard core. The opposite is observed in the red curve of the figure~\ref{fig:coarse-grained-RDF}, correspondent to NP with polymers free to rotate, where the cores can approximate easily, increasing the occupancy in the first length scale and decreasing in the second length scale.

From these RDFs curves, using both the solution of the Ornstein-Zernike equation~\cite{head-gordon92} with integral equation approximation,  the effective potentials for polymer-grafted nanoparticles with fixed and free beads  were obtained~\cite{head-gordon92, yan08, Lafitte14}. The potentials were also obtained using the inverse-Boltzmann procedure~\cite{Alamarza04}. Essentially the same potentials  were obtained by both methods.  

\begin{figure}[htb]
	\centering  \includegraphics[width=6cm]{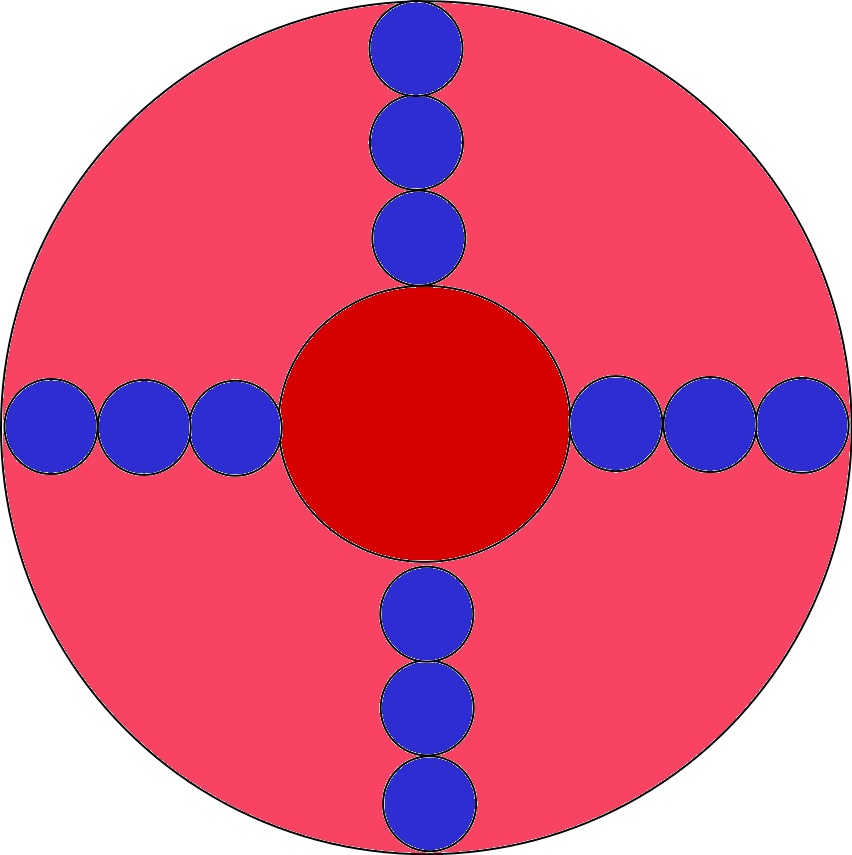}
	
\caption{Schematic depiction of the effective nanoparticles. It has a central hard core (the red sphere) and a soft corona (the lighter red 
sphere).}
\label{fig:core-softened-particles}
\end{figure}

The polymer-grafted nanoparticles become represented by spherical particles interacting through these  effective core-softened potentials as illustrated in figure~\ref{fig:core-softened-particles}. 

\begin{figure}[htb]
\centering \includegraphics[scale=0.6]{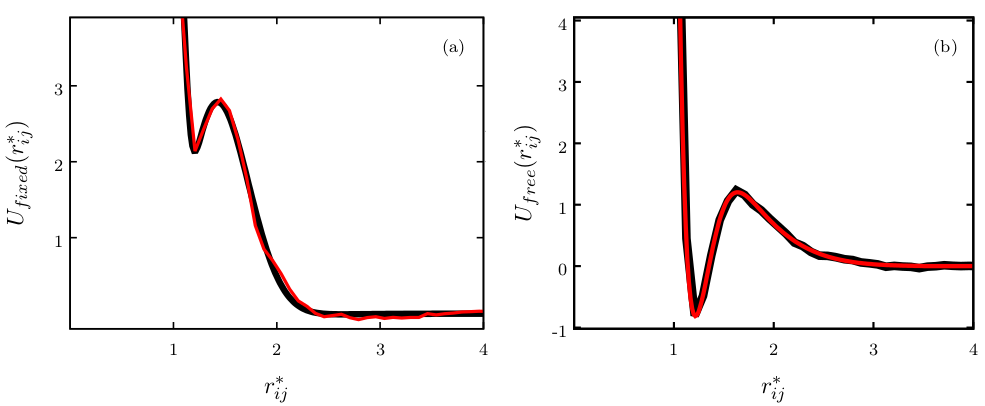}
\caption{Core-softened potential for polymer-grafted nanoparticles  with four monomers   (a) fixed and (b) free to rotate around nanoparticle core. The red
curve is the potential obtained by solving the Ornstein-Zernike equation, and the black curve is the LJ+Gaussian fit.}
\label{potentials}
\end{figure}

Based in previous works~\cite{head-gordon92, yan08, Ney2009}, our effective potentials are composed by a short-range attractive 
Lennard Jones potential and three Gaussian terms,  
each one centered in $c_j$, with depth $h_j$ and width $w_j$:
\begin{equation}
U(r_{ij})=4\epsilon_{core} \left [\left (\frac{\sigma_{core}}{r_{ij}}  \right )^{12} - \left (\frac{\sigma_{core}}{r_{ij}}  
\right )^{6}  \right ]+ \sum_{j=1}^{3}h_jexp\left [ -\left ( \frac{r_{ij}-c_j}{w_j} \right )^2 \right ], \; .
\end{equation}
\noindent Here, $r_{ij} = |\vec r_i - \vec r_j|$ is the distance between two cores $i$ and $j$. The resulting potentials and fittings
are shown in the figure~\ref{potentials} for case of  NP with polymers fixed ($U_{fixed}$) or free ($U_{free}$) to rotate.  
The parameters correspondent to each case are given in the table \ref{table1}.

\begin{table}
\centering
\setlength{\tabcolsep}{12pt}
\begin{tabular}{c c | c c}
\hline
\multicolumn{2}{c|}{$\mathbf{U_{fixed}}$ \bf potential}  & \multicolumn{2}{|c}{$\mathbf{U_{free}}$ \bf potential} \\ 
\hline
Parameter & Value & Parameter & Value \\ 
\hline 
$h_1$ & 0.287379 & $h_1$ & -3.80084 \\ 
$c_1$ & 2.055295 & $c_1$ & 1.11192 \\ 
$w_1$ & 1.526922 & $w_1$ & 0.313324 \\ 
$h_2$ & 2.706034 & $h_2$  & 46.1324 \\ 
$c_2$ & 1.461441 & $c_2$  & 0.774361 \\ 
$w_2$ & 0.37436409 & $w_2$  & 0.191852 \\ 
$h_3$ & -0.0650439 & $h_3$  & 6.37621 \\ 
$w_3$ & 2.9884193 & $w_3$  & 0.192937 \\ 
$c_3$ & -0.4264772 & $c_3$  & 1.23615 \\ 
\hline
\end{tabular}
\caption{Parameters of the particle-particle potentials in reduced units.}
\label{table1}
\end{table}

In the effective potentials is also clear the effect of the polymers mobility. When they are held fixed the energetic penalty 
for two NPs move from the further (or second) scale to the closer (or first) scale is higher than in the case when the polymers can 
rotate and expose one core to another. As consequence, the $U_{fixed}$ potential has a ramp-like shape, while the  $U_{free}$ a short range attraction 
and a long range repulsion (SALR) shape.

\subsection{Simulation Details for the Effective Potential}

The systems consists of 800 disks with diameter $\sigma = \sigma_{core}$.  Langevin Dynamics simulations 
were performed with a time step of $\delta t = 0.001$. Periodic boundary conditions were applied in both directions. 
We performed $5\times10^5$ steps to equilibrate the system. These steps were then followed by $2\times10^6$  steps for 
the results production stage. To ensure that the system was thermalized, the pressure, kinetic and potential energy 
were analyzed as function of time. The velocity-verlet algorithm was employed to integrate the equations of motion. 
The $NVE$ ensemble was employed for equilibration and $NPT$ ensemble for the production runs.  
The Langevin thermostat, with a damping parameter $\gamma = 1.0$, was employed to fixed the system 
temperature and the pressure was held fixed by the Nos\`e-Hoover barostat with a parameter 10$\delta t$. 
The simulations of the effective model were performed using the Large-scale Atomic/Molecular Massively Parallel Simulator 
(LAMMPS) package~\cite{plimpton-jcp1995} and the $PT$, $T\rho$ and $P\rho$ phase diagrams for each potential were obtained.

The dynamic anomaly was analyzed by the relation between the mean square displacement (MSD) and time, namely
\begin{equation}
\left \langle [r(t)-r(t_0)]^2 \right \rangle = \left \langle \Delta r^2(t) \right \rangle,
\end{equation}
where $r(t_0) = (x^2(t_0) + y^2(t_0))^{1/2}$ and $r(t) = (x^2(t) + y^2(t))^{1/2}$ denote the coordinate of 
the particle at a time $t_0$ and at a later time t, respectively. The MSD is related to the diffusion coefficient $D$ by\cite{allen2017}
\begin{equation}
D = \lim_{t\rightarrow \infty } \frac{\left \langle \Delta r^2(t) \right \rangle}{4t}.
\end{equation}
The structure of the fluid was analyzed using the radial distribution function (RDF) $g(r_{ij})$.  In order to 
check if the system exhibits the density anomaly, the temperature of  maximum density (TMD) was computed for different isobars in $T\rho$ diagram.

The phase boundaries were defined analyzing the specific heat at constant pressure, $C_P$~\cite{allen2017},the system mean square displacement, 
the radial distribution function and the discontinuities in the density-pressure phase diagram.

\section{Results and discussion}

Here we analyze the thermodynamic and dynamic behavior  of the system of polymer-grafted nanoparticles 
represented by the effective core-softened potentials generated for fixed and  free beads systems. 

\subsection{Polymer-grafted nanoparticles with fixed polymers}

\begin{figure}[htb]
	\centering \includegraphics[scale=0.6]{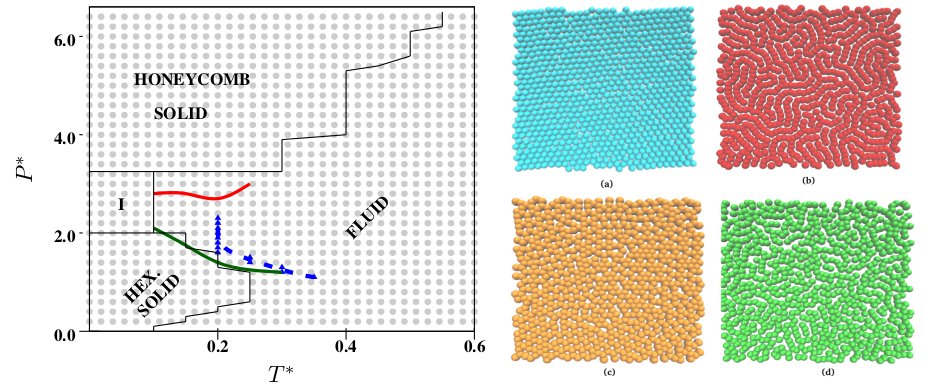}
\vspace{-1\baselineskip}
\caption{left panel: Pressure versus temperature phase diagram of system with fixed polymers.  The gray lines are the isochores,the black lines divide the distinct phases, I represents the hexagonal solid phase, the blue line indicates the TMD,  the green and red lines are  the maxima and minima in diffusion coefficient. Right panel:  System snapshots for: (a) hexagonal solid($P^*=0.40$ and $T^*=0.05$); 
(b) amorphous solid($P^*=2.80$ and $T^*=0.10$); (c) honeycomb solid($P^*=5.00$ and $T^*=0.20$); (d) fluid ($P^*=5.00$ and $T^*=0.80$).}
\label{diagramPT_Uno}
\end{figure}

\begin{figure} [htb]
\includegraphics[scale=0.6]{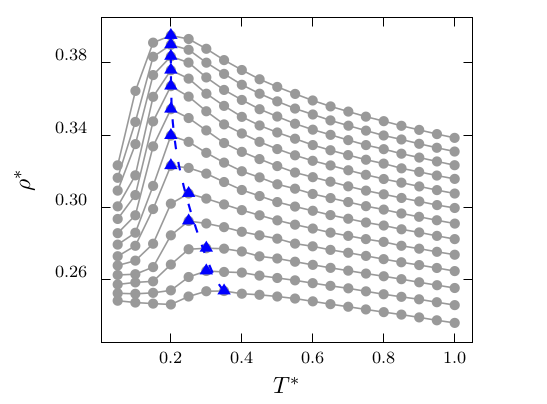}
\caption{Temperature of Maxima Density for isobars between $P^*=1.10$ and $P^*=2.30$ from bottom to top.}
\label{fig3}
\end{figure}

The pressure versus temperature  phase diagram obtained using the effective potential for the grafted nanoparticles with 
fixed polymers (see the potential in figure~\ref{potentials}(a)) is illustrated in figure~\ref{diagramPT_Uno}. 
Three solid structures were observed. At lower pressures, a hexagonal solid was obtained, as shown in the 
snapshot~\ref{diagramPT_Uno}(a). Increasing the pressure  the system enters in the region were the anomalous behavior is  observed - 
the waterlike anomalies which will be discussed next. A consequence of the anomalies in the phase diagram is the presence of a 
reentrant liquid phase and a transition from the well defined hexagonal lattice to an amorphous stripe-like structure. This 
ordered-disordered transition was observed in previous works where particles interact through two length potentials also known as the   
ramp-like potentials~\cite{Bok16a, Bok16b, Bordin18b}. In the previous works as here the anomalies arise from the competition between the two length scales. The figure~\ref{fig3} illustrated the density versus temperature for fixed pressure showing the maximum density, a water-like anomaly.

\begin{figure}[htb]
\begin{center}
\includegraphics[width=08cm]{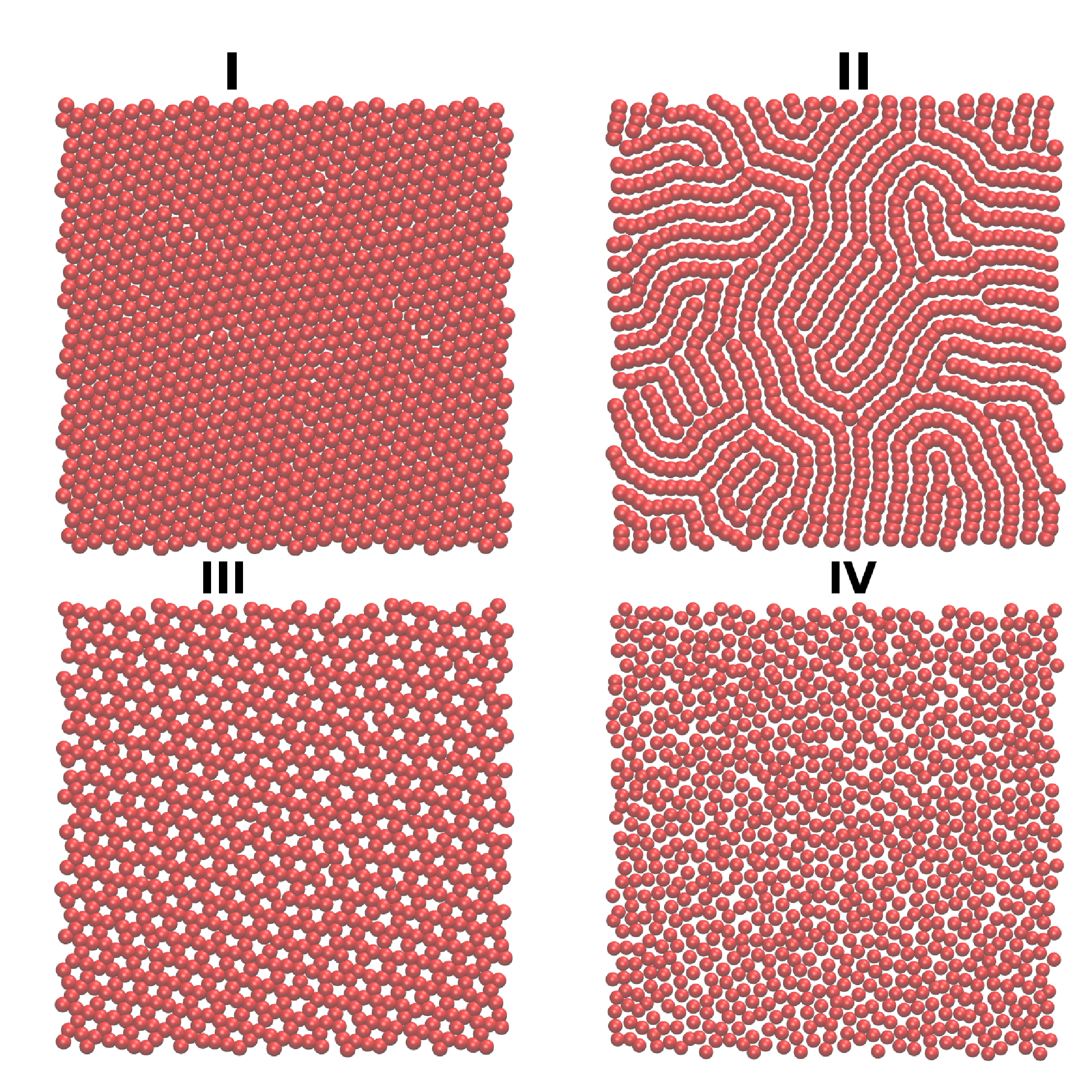}
\end{center}
\caption{Comparison between the patterns observed in the CG model and effect model.}
\label{snapCGmodelfixed}
\end{figure}

The effective model is obtained from the coarse-grained system using a radial distribution function  for one specific temperature and pressure. This raises the question of how reliable is this approach to describe the system for many pressures and temperatures. 
In order to test how robust is the effective model, we performed additional simulations for the coarse-grained description of the polymer-grafted colloidal system in the region where the anomalous behavior in the effective model was observed. Then, new  $NPT$ simulations of the CG  system composed of 1000 NPs. Four points in the phase diagram were selected: (I) $T^* = $ 0.10 and $P^*$ = 1.0 (inside the hexagonal solid region of the effective model phase diagram), (II) $T^* = $ 0.10 and $P^*$ = 3.0 (inside the stripe solid region of the effective model phase diagram), (III) $T^* = $ 0.10 and $P^*$ = 4.0 (inside the honeycomb solid region of the effective model phase diagram) and  (IV) $T^* = $ 0.20 and $P^*$ = 2.0 (inside the reentrant fluid phase of the effective model phase diagram). Figure~\ref{snapCGmodelfixed} illustrates these state points. The structures are similar to the obtained using the effective model, figure~\ref{diagramPT_Uno}. This indicates that the effective model was able to capture the proper behavior of the CG model phase diagram.

One of the characteristics of systems interacting through two length scales
potential as the potentials illustrated in the figure~\ref{potentials} is 
the presence of thermodynamic anomalies. The density anomaly is characterized by a maximum in the $\rho(T)$ curve along a isobar. 
For constant pressure as the temperature increases the density 
increases by making particles to rearrange from one length scale to the
other. This can be also observed in the 
radial distribution function  $g(r_{ij})$ which 
presents  two peaks: one
at the closest scale, $r_1$, and another at the furthest scale, $r_2$.
Recently it has been suggested that a signature of the presence of TMD line would be
given by the radial distribution function as follows: at fixed temperature, as the density is
increased, the radial distribution function of the closest scale, g($r_1$), would increase its value, while the 
radial distribution function of the furthest scale, g($r_2$), would decrease~\cite{evy2013}.  This can also be represented by
the rule~\cite{franzese2010,Raposo2014}:
\begin{equation}
\Pi _{12}=\frac{\partial g(r) }{\partial \rho}\mid_{\rho_1}  \times \frac{\partial g(r) }{\partial \rho}\mid_{\rho_2}<0 \;.
\label{competition_tmd}
\end{equation}

\begin{figure}[htb]
        \centering \includegraphics[scale=0.7]{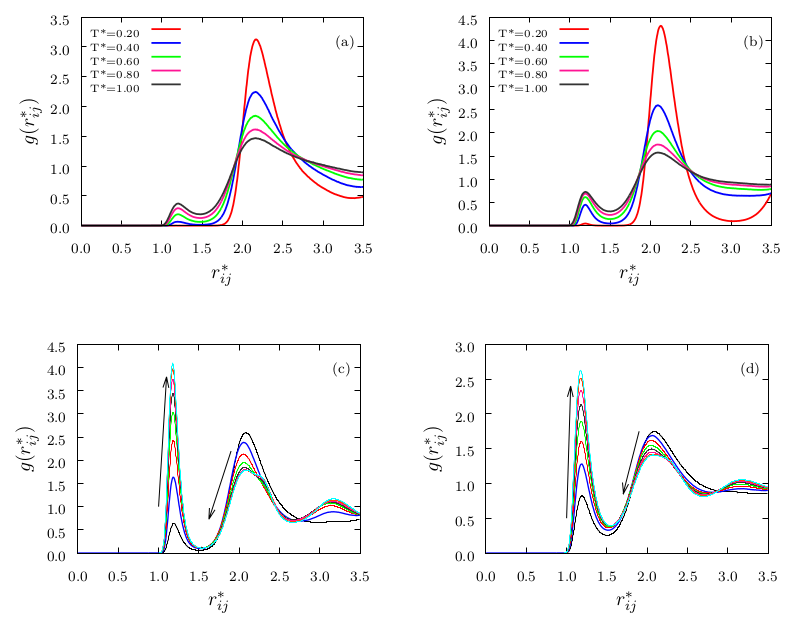}
\caption{Radial distribution function behaviour for pressure two values below the threshold and temperature variation ($P^*=0.40$ in (a) and $P^*=0.80$ in (b)) indicating thermal effects in this region. At bottom figures, RDF's for two fixed temperatures ($T^*=0.40$ in (c) and $T^*=0.80$ in (d)) and pressure variation inside range $0.90<P^*<3.00$ indicating pressure effects in TMD region.}
\label{fig:rdfs_termal_effects}
\end{figure}

The physical picture behind this condition is that, for a fixed pressure, as the temperature increases, particles that are located at the attractive scale, $r_2$, move to the repulsive scale, $r_1$ - the thermal effects, which occur up to a certain pressure threshold $P_{min}^*=0.90$. For pressures in range $0.90<P^*<3.00$, for a fixed temperature, as the pressure increases, particles exhibit the same offset between the potential length scales $r_1$ and $r_2$ - the pressure effects. Figure \ref{fig:rdfs_termal_effects} illustrate a typical radial distribution functions at fixed $T^*$ as $P^*$ is varied [(a) and (b)] and vice-verse [(c) and (d)].

The regions identified by the radial distribution function as fulfilling the condition Eq. \ref{competition_tmd} are illustrated as red circles in figure \ref{tmd_and_rdf2}(a). The solid curve shows the TMD line. All the stable state points with density equal or higher the minimum density at the TMD line verify the relation $\Pi _{12}(\rho, T) < 0$. This result gives support to our assumption that the presence of anomalies is related to particles moving from the furthest scale, $r_2$, to closest length scale $r_1$. In addition it indicates that the two length scales in the effective potential are related to the core-core repulsion competing withe polymer-polymer attraction present in the coarse-grained potential.

\begin{figure}[htb]
		\centering \includegraphics[scale=0.6]{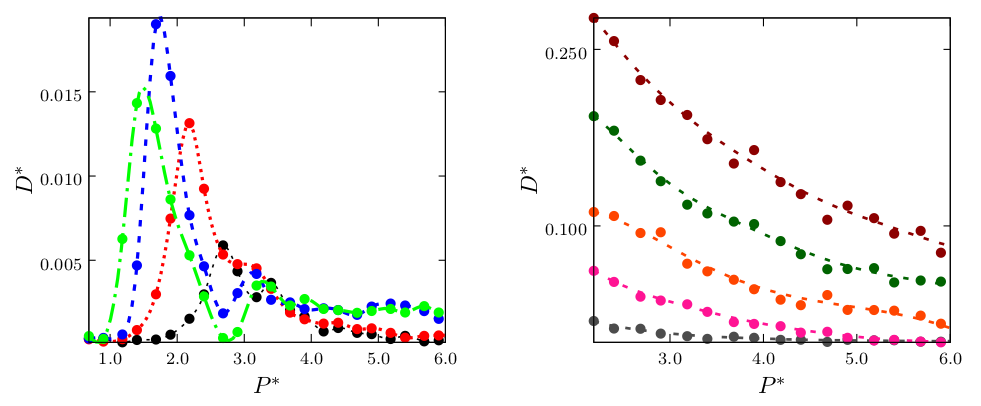}
\caption{Diffusion coefficient versus pressure for  (a) $T^*=0.05$ (black line), $T^*=0.10$ (red line), $T^*=0.15$ (blue line), 
$T^*=0.20$ (green line) and (b) $T^*=0.30$ (gray line), $T^*=0.40$ (magenta line), $T^*=0.50$ (orange line), $T^*=0.60$ (green line), $T^*=0.70$ (brown line).}
\label{diffusion_Uno}
\end{figure}

\begin{figure}[htb]
		\centering \includegraphics[scale=0.6]{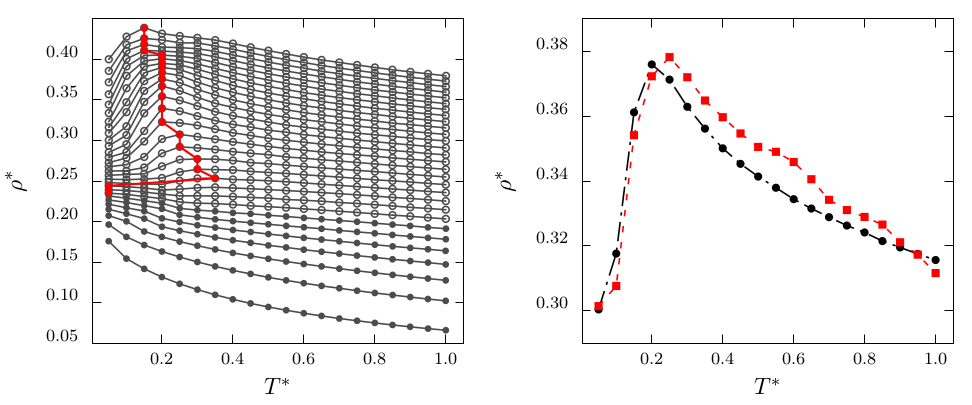}
\caption{(a) TMD line (red line) for distinct isobars in the effective model (b) Comparison of one $\rho(T)$ curve along the isobar $P^* = 2.0$ between the effective (black circles) and the CG model (red squares).
}\label{tmd_and_rdf2}
\end{figure}

Another signature of anomalous fluids is the behavior of the the diffusion coefficient which increases with density. Figure \ref{diffusion_Uno} represents the diffusion coefficient versus pressure for different  isotherms, showing  that  D in a certain range of temperatures and pressures increases with pressure. The minimum in the diffusion coincides with the melting line. This behavior of the diffusion and melting line is related to  ordered - disordered transition  and it was previously observed for  ramp-like potentials in two dimensions~\cite{Bordin18b}. 

Finally in order to check if the CG model also shows anomaly, we run simulations along the isobar $P^*  =2.0$. Figure~\ref{tmd_and_rdf2} (b) illustrates the density versus temperature for  $P^*  =2.0$ for both CG (red squares) and 
effective (black circles) potentials. The two behaviors are quite similar. This result indicates that our strategy to derive 
a simpler two length scales potential to describe a more sophisticated system obtaining some information about the origin of the anomaly is valid.

\subsection{Polymer-Grafted nanoparticles with free nanoparticles}

\begin{figure}[htb]
		\centering  \includegraphics[scale=0.6]{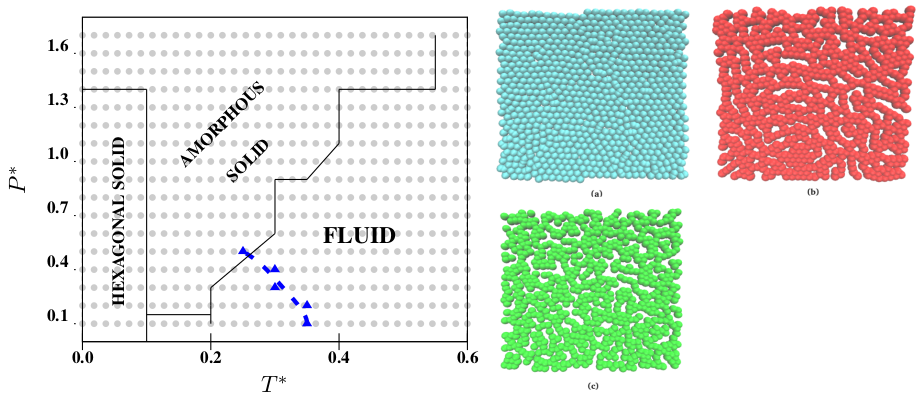}
		\caption{left panel: Pressure temperature phase diagram of system which polymers are free to rotate. 
The gray dots are the simulated points. The lines divides the distinct phases, I is the hexagonal phase, 
 II is the amorphous solid and  the  blue line is the  TMD. right panel: System snapshots for 
(a) hexagonal solid ($P^*=0.40$ and $T^*=0.05$); (b) amorphous solid 
($P^*=1.10$ and $T^*=0.10$) and (c) fluid ($P^*=1.10$ and $T^*=0.70$).}
\label{diagramPT_Uyes}
\end{figure}

The pressure versus temperature  phase diagram obtained using the effective potential for the grafted nanoparticles with free polymers is illustrated in figure~\ref{diagramPT_Uyes}. The phase behavior of the system is quite distinct when compared with the phase diagram for the system with fixed polymers.

At low temperatures ($T^*\le 0.10$) , and for pressures up to $P^*=1.40$, the system is in a hexagonal solid phase. Increasing the temperature for $P^* < 0.12$, the system melts to a fluid phase, while in the range $0.12< P^* < 1.4$ there is a order-disordered transition in the solid structure, that changes from the hexagonal to amorphous.

Both free and fixed polymers systems show a number of similarities in the phase space, here, however,  we do not observe a reentrant fluid phase neither the honeycomb solid phase. Also, the solid-liquid separation line moves to higher temperatures. As consequence, the TMD line is smaller and no diffusion anomaly is present.

The absence of diffusion anomaly when the system exhibits a TMD is not new. It arises in lattice systems in the presence of two length scales interactions depending of the balance between the two length scales~\cite{Szortyka09} and in confinement due the competition between the length scales and the confinement~\cite{BoK17}. In this case, it may be related to the fact that here there is no reentrant fluid region - as we saw, this two phenomena were correlated for the $U_{fixed}$ potential.

\begin{figure}[htb]
\centering \includegraphics[scale=0.6]{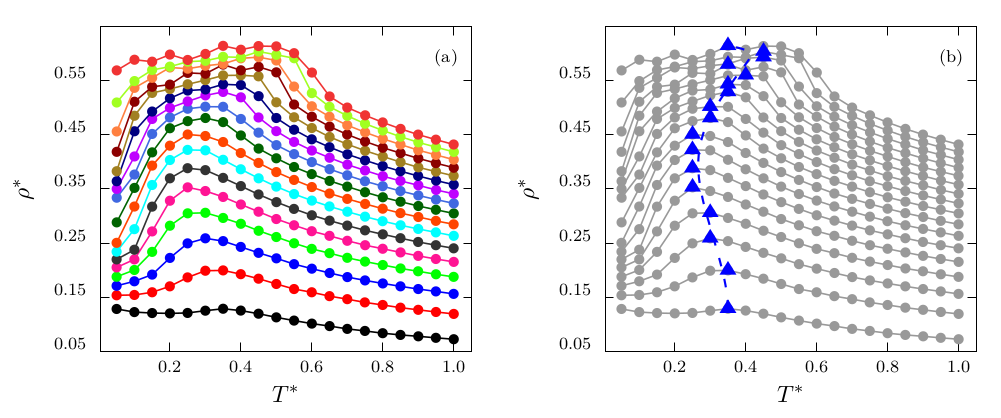}
\caption{Density-temperature phase diagrams. (a) The color lines are the isobars from $P^*=0.10$ (bottom) 
to $P^*=1.70$(top). (b) The same diagram explaining the maximum density temperature (TMD) and the peaks in 
the specific heat at constant pressure (which coincide with the inflection points of the isobars with $P^*<1.00$.}\label{fig:lab}
\end{figure}

Usually the presence of the  TMD, as  shown in figure~\ref{fig:lab}, is related to a competition between the two length scales, as discussed earlier. However, some studies has shown that this same phenomena can occur in fluids without competitive scales, but just a weak softening of the interparticle repulsion can lead to anomalous behavior.~\cite{Prestipino10, Prestipino11}.

\begin{figure}[htb]
        \centering \includegraphics[scale=0.7]{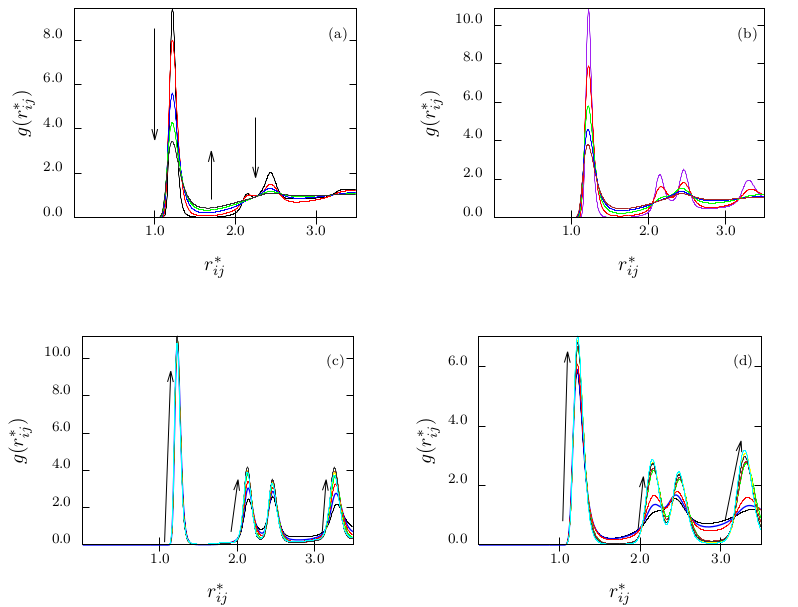}
\caption{Radial distribution function behaviour for maintaining temperature fixed and varying pressure and 
vice-verse in order to verify relation between competition and structure. (a) $P^*=0.40$ and each curve is 
from one temperature. (b)$P^*=0.80$  analogously. (c) $T^*$ is fixed at $T^*=0.20$ and pressure is varying. (d)$T^*$ is fixed at $T^*=0.60$, analogously}
\label{fig:rdfs_termal_Uyes}
\end{figure}

\begin{figure}[htb]
\centering \includegraphics[scale=0.7]{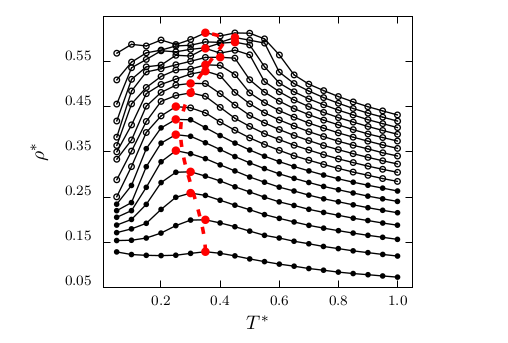}
\caption{Density - temperature phase diagram confirming that's no relation between competition and structure. 
Filled symbols are dominated by thermal effects, and empty are those dominated by pressure effects. }
\label{tmd_and_competition}
\end{figure}

Therefore, unlike the previous case ($U_{fixed}$), it is not possible to establish the connection between structure and anomaly in density, as we can see in figure \ref{fig:rdfs_termal_Uyes}, which shows behaviour of RDF's by varying temperature (at fixed $P^*$) and pressure (at fixed $T^*$). This disconnection can also be analyzed taking into account that the unfilled points of the graph obey the relation between the migration of scales, and that , in turn, the TMD reaches all points (filled or not), it is concluded that, for this potential, it would not be the competition between the scales responsible for the density anomaly, as may seen in figure \ref{tmd_and_competition}.

\section{Summary and conclusions}
In this work, a two dimensional system of polymer grafted nanoparticles is analyzed using large-scale Langevin
Dymanics simulations. The use of effective core-softened potentials allow us to explore the complete system phase space. 
In this way, the $PT$, $T\rho$ and $P\rho$ phase diagrams for each potential were obtained The
phase boundaries were defined analyzing the specific heat at constant pressure, the system mean square displacement, the 
radial distribution function and the discontinuities in the density-pressure phase diagram. Also, due the competition in 
the system we have observed the presence of water-like anomalies, such as the temperature of maximum density - in addition
with a tendency of the TMD to move to lower temperatures (negative slope)- and the diffusion anomaly. It was observed  
different structural morphologies  for each nanoparticle case. We observed that for the fixed 
polymers case the waterlike anomalies are originated by the competition between the potential characteristic length scales,
while for the free to rotate case the anomalies arises due a smaller region of stability in the phase diagram and no competiton
between the scales was observed.

The main driving force for these different morphologies obtained is the competition between strong short-range attractions of the particle cores (the enthalpic gain upon the core-core aggregation) and long-range entropic repulsions of the grafted chains. 

\section{Acknowledgments}

The authors thank the financial support from the Brazilian agencies FAPERGS and CNPq. TPON thanks the Coordena\c{c}\~ao de Aperfei\c{c}oamento de Pessoal de N\'ivel Superior (CAPES), Finance Code 001. The authors thank Prof. Alan Barros from Universidade Federal de Ouro Preto, Minas Gerais, Brazil, and Prof. Enrique Lomba from Instituto de Quimica Fisica Rocasolano, Madri, Spain, for computational time in cluster to run the effective model simulations.

\bibliography{biblio}

\end{document}